\documentclass[aps,preprint]{revtex4}
\usepackage{amsfonts}
\usepackage{amsmath}
\usepackage{amssymb}
\usepackage{graphicx}
\usepackage{bm}
\usepackage{color}

\begin{document}

\title{Dynamical Origin of Power Spectra}
\author{T. Kodama}
\affiliation{Instituto de F\'{\i}sica, Universidade Federal do Rio de Janeiro, C. P.
68528, 21945-970, Rio de Janeiro, Brasil}
\author{T. Koide}
\affiliation{Institut f\"ur Theoretische Physik, Johann Wolfgang Goethe-Universit\"at,
60438, Frankfurt am Main, Germany}
\pacs{PACS number}

\begin{abstract}
We discuss a possible origin of Tsallis' statistics from the correlation
among constituents which reduces the phase space of the system. We also show
that in the system of coupled linear harmonic oscillators can exhibit a
Tsallis type behavior.
\end{abstract}

\maketitle

\section{Introduction}

If we want to determine precisely the dynamical behavior of a macroscopic
system out of equilibrium, we need in general to deal with highly coupled
equations of motion related to the full microscopic degrees of freedom of
the system. However, when we are interested in knowing only the dynamics of
macroscopic observables, we can sometimes expect a large amount of reduction
of degrees of freedom relevant to the dynamics of these macroscopic
observables. They should be described by a closed set of simpler
coarse-grained equations. Such a situation happens when the space-time
scales for our macroscopic observables are clearly separated from those of
the microscopic ones. The hydrodynamics is such an example.

Hydrodynamics is derived by assuming the local thermal equilibrium: at any
point of a fluid, there exists a \textit{finite} domain (fluid cell) in
which the fluid can be regarded as homogeneous and in thermal equilibrium,
so that the usual thermodynamic relations hold for this piece of fluid. The
physical properties of the fluid are determined by these thermodynamic
relations. Furthermore, although finite, the size of the domain is
considered to be negligible compared to the typical scale of the system, so
that the thermodynamic variables can be considered as continuously varying
functions of position and time.

On the other hand, strictly speaking, the theromodynamic description of a
matter is only possible when the time scale in question is much larger than
the relaxation time of the matter needed to attain thermodynamic
equilibrium. In other words, for hydrodynamic description, the fluid cell
should have the minimum scale to realize local thermal equilibrium. 
For example, for a gas, such a domain should at least be several times
larger than the mean-free path of the gas particles, otherwise fluctuations
are too large and the thermodynamic description of the gas property will
fail. It is thus reasonable to consider that hydrodynamic approach is not
applicable when the size of the system is relatively small so that the
typical space-time inhomogeneity scales are comparable or smaller than those
required for the local thermalization of the system. In fact, it is believed
that the deviations from the usual thermodynamics is an important factor in
physics associated with molecular scales such as biological systems \cite%
{small}.

A parameter commonly used to evaluate quantitatively the applicability of
hydrodynamic approach is the Knudsen number $K$, defined as the ratio, $%
K=\lambda _{meso}/\lambda _{hydro}$ where $\lambda _{meso}$ is the
mesoscopic scale of the microscopic structure (in the case of a gas, it is
the mean-free path of the gas particles ) and $\lambda _{hydro}$ is the
typical hydrodynamical inhomogeneity scale. If $K\ll 1$ everywhere, a
hydrodynamic description will be applicable as the dynamics of the system.
This also means that the time scale of the macroscopic (hydrodynamic) scales
are much larger than the thermodynamic relaxation time.\textbf{\ }In
general, the hydrodynamic time scale is of the order of $\lambda
_{hydro}/v_{sound}$ , where $v_{sound}$ is the velocity of the sound of the
fluid.

On the other hand, there are several counter examples where hydrodynamic
models seem to work for $K$ values apparently not so small. One of these is
the physics of relativistic heavy-ion collisions \cite{hydro}. To analyse
the behaviors of huge amount of produced particles in the relativistic
heavy-ion collisions, we simply assume that the many-body dynamics can be
approximately given by relativistic hydrodynamics (in particular, perfect
fluid). Since the typical dimension of the whole system is initially from few%
$\ fm$ ($10^{-15}m$) (or even less if we consider the Lorentz contraction
due to the relativistic effects) and expands violently. For a hydrodynamic
description to be valid from the beginning of this expansion, we expect that
the fluid cell size should at least be less than one order of magnitude of $%
1fm$. That is, we should use the cell size of the order of $0.1fm$, which is
already comparable, if not smaller, to a typical value of mean-free path of
the constituent particles, partons. The value, $0.1fm$ is even closer to the
scale of quantum correlation length. In spite of these facts, the collective
motion of the matter produced in the relativistic heavy ion collision is
well described by the hydrodynamic model. The hydrodynamic model is a very
important tool to understand the experimental data.

Then how should we understand such a situation? Here, we have to recall that
the local thermodynamic equilibrium is a \textit{sufficient condition} for
the thermodynamic relations to be valid, but not \textit{a necessary
condition}. For example, let us consider an ideal Boltzmann gas. If the
system is isotropic in momentum space at any position, the energy density
and pressure are given as%
\begin{equation}
\varepsilon \left( \vec{r}\right) =\int d^{3}p\ E\left( \vec{p}\right)
f\left( \vec{p},\vec{r}\right) ,
\end{equation}%
and%
\begin{equation}
P\left( \vec{r}\right) =\frac{1}{3}\int d^{3}p\ \left( \vec{p}\cdot \vec{v}%
\right) \ f\left( \vec{p},\vec{r}\right) ,
\end{equation}%
where $f\left( \vec{p},\vec{r}\right) $ is the one-particle phase space
distribution function and $\vec{v}=\vec{p}/E$. If these particles are
massless (or ultra-relativistic), then $\left\vert \vec{p}\right\vert
=E\left( \vec{p}\right) ,$ so that we always have\footnote{%
In this paper, we use the natural unit, $c=\hbar =1.$}%
\begin{equation}
\varepsilon =3P,  \label{e=3p}
\end{equation}%
independent of the form of the distribution function $f$ \footnote{%
More generally, Eq.(\ref{e=3p}) is valid if the four-energy momentum tensor
is traceless (conformal theory) and isotropic in the Landau local rest
frame. For a hydrodynamic description, we only need the thermodynamic
relations (equation of state) as Eq.(\ref{e=3p}).}.

As we saw above, what we need to arrive Eq.(\ref{e=3p}) is only the
condition of\textit{\ local isotropization }of the particles in the momentum
space, and not the\textit{\ thermal equilibrium}. It may well be possible
that the process of local isotropization in momentum space can be attained
much faster than the complete thermalization. In fact, for a quark-gluon
plasma which is now believed to be produced by relativistic heavy-ion
collisions, such a dynamical mechanism has been studied in analogy with the
plasma instabilities \cite{Weibel}. This is an example where the
thermodynamic relations become valid very quickly even the system is not in
a real thermal equilibrium. It is also shown that the nonequilibrium
dynamics of quantum scalar field exhibits such a prethermalization behavior 
\cite{berges}. These considerations indicate that there are situations where
hydrodynamical description can still be valid even for a system where the
usual criteria in terms of Knudsen number is apparently violated. From now
on, we refer such a state of matter to \textquotedblleft
prethermalized\textquotedblright\ state, for which functional relations
among macroscopic observables (such as equation of state) emerge, although
the true thermalization is not yet attained.

Now, two basic questions arise here. One is in which situations such a
prethermalized state occur, and the second, whether the thermodynamics
properties (equation of sates) of the matter stay the same as
those of the real thermal equilibrium, or not. Either prethermalization or
real thermalization, the emergence of functional relations among macroscopic
observables indicates the reduction of the other microscopic degrees of
freedom irrelevant to the description of the physical properties of the
matter. This must be due to a kind of randomization processes of these
irrelevant degrees of freedom (irrelevant variables) where the information
of the initial condition is lost by random interactions among them. The
difference between the real thermalization and pre-thermalization is that,
in the case of pre-thermalization, there still exist some degrees of freedom
which still depends on the initial condition.

If such a prethermalization scenario happens, we can expect some universal
behavior such as the Boltzmann-Gibbs statistics which leads to the
thermodynamic relations. As a matter of fact, the Boltzmann-Gibbs statistics
is not a unique possibility to be consistent with thermodynamics. In 1988,
Tsallis proposed a new statistical mechanics where entropy is non-additive 
\cite{tsallis,tsallis2}. So far, extensive studies have been
developed and the applications of Tsallis' entropy extend to biological
systems, high-energy physics, and cosmology. Because of the non-additivity
of the entropy, Tsallis statistics is considered to be important for
describing correlated systems. From our prethermalization scenario, the
Tsallis statistics may be attained before the real thermalization is
achieved and the later relaxation process after the Tsallis statistics may
be described by the time evolution of $q$, which is the parameter of Tsallis
statistics, and the Tsallis statistics is reduced to the Boltzmann-Gibbs
statistics for $q=1$ \cite{kodama,kodama2}.

To make the idea clear on the effect of correlation on the relaxation time
of a prethermalization mechanism, let us consider a problem of finding a
minimum of a function of two variables, as shown in Fig. 1. When the
function has a narrow long valley, starting from some arbitrary initial
point, the usual steepest descent method always first goes to the closest
point of the valley instead of going directly to the real minimum, as
indicated by the dashed curve. The narrow valley means that there exists a
strong correlation between the two variables. In other words, if there exist
any strong correlations among variables, the minimizing path always tries to
satisfy the correlation first (fast relaxation stage), then follows
downstream to the true minimum along the valley generated by the correlation
(slow relaxation stage).

\begin{figure}[tbp]
\includegraphics[scale=0.6]{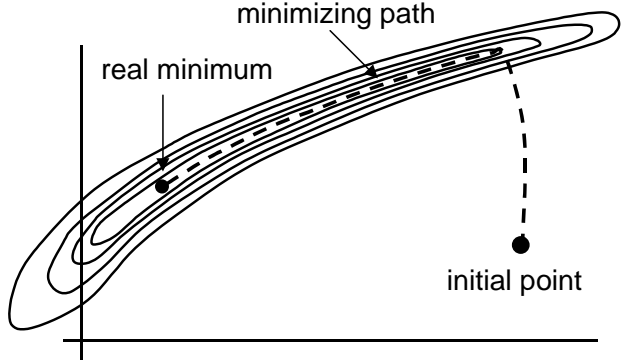}
\caption{Presence of strong correlation among the two variables in
optimisation problem.}
\label{fig1}
\end{figure}

The above consideration suggests that a prethermal state may correspond to a
kind of narrow valley of the free energy generated by dynamical
correlations. If such a local minimum stays for a very long time, we may
consider the system as in a kind of equilibrium, but the microscopic
occupation probability is not that of the Boltzmann-Gibbs statistics, due to
the constraint created by the correlation. Then what characterizes such a
local equilibrium? We show that there exists a class of dynamical
correlation for which the minimisation of the energy under the constraints
lead exactly to the general form of Tsallis' statistics. In this image, the
dynamics of the slow relaxation stage may be described by the evolution of $%
q $ to arrive at the real thermal equilibrium described by the
Boltzmann-Gibbs statistics corresponds to $q=1$.

Note that in the above analogy, the correlation is rather static one.
However the dynamical constraints are not necessarily determined only from
the Hamiltonian itself, but may depend on the state of the system in a
dynamical way. In other words, for a given system, depending on the initial
condition, the correlation and hence the constraints among microscopic
degrees of freedom may change. In this sense, it is possible that the value
of $q$ is not a quantity determined from the beginning even when the system
is specified.

Of course, prethermalization involves a broad range of phenomena, and has
been studied for a long time in several works \cite{jou}. In particular, the
relaxation to a quasi-stationary state under long-range interactions is
strongly suggested in the Hamiltonian mean field model \cite{hmf}. Here we
develop a general phenomenological argument of how the states described by
the Tsallis nonextensive statistics can be regarded as a prethermalized
states.

\section{Dynamical Clusters and Tsallis Distribution Function}

In this section, we show that the Tsallis distribution function can be
obtained when there exists a class of correlations among particles in each
single particle energy levels.

Let us consider a system composed of $N$ particles where the single-particle
energy spectrum and the corresponding occupation numbers are given by $%
\left\{ \varepsilon_{\alpha}\right\} $ and $\left\{ N_{\alpha}\right\} $,
respectively.

Suppose there exists a strong $q$-body correlation among $N_{\alpha}$
particles in any single-particle state $\alpha$. In general, $q$ is a
function of $\alpha$. However, we assume, for simplicity, the number of
correlated particles are independent of state $\alpha$. Due to this
correlation among particles, the rules to determine the occupation numbers $%
\left\{ N_{\alpha }\right\} $ affects the statistics as described below. For
a given value of $q$, the number of possible ways of forming correlated $q$%
-body clusters in each $\alpha$ state is given by 
\begin{equation}
\left( 
\begin{array}{c}
N_{\alpha} \\ 
q%
\end{array}
\right) =\frac{N_{\alpha}!}{q!\left( N_{\alpha}-q\right) !}\approx \frac{%
N_{\alpha}^{q}}{\Gamma(q+1)},  \label{M_alfa}
\end{equation}
for $N_{\alpha}\gg1$. Here, we used Stirling's formula for the asymptotic
values of $\Gamma$-function. Note that $q$ is the average number of
particles to form a cluster and does not have to be integer. We call these
correlated subsystems $q$-clusters. If the $q$-clusters are formed with
equal a priori probability, the number of such clusters $M_{\alpha}$ should
be proportional to the number of ways of forming them. Thus%
\begin{equation}
M_{\alpha}=Const\times N_{\alpha}^{q},  \label{Ma}
\end{equation}
and the total number of the $q$-clusters in the whole system is then 
\begin{equation}
M=\sum_{\alpha}M_{\alpha}.
\end{equation}

We may estimate the energy associated with the cluster formation. Since a $q$%
-cluster in the energy level $\alpha$ carries the single particle energy, $%
q\times\varepsilon_{\alpha},$ we have 
\begin{equation}
E_{q}=q\sum_{\alpha}M_{\alpha}\varepsilon_{\alpha}.  \label{E_q}
\end{equation}
Note that, for $q\rightarrow1,$ the expression above formally leads to the
total energy of the system, except for a constant factor This fact will be
useful for the recovering the usual Boltzmann-Gibbs distribution. See the
later discussion.

It is natural to think that, once the number of the correlated particles are
fixed by $q$, the most probable configuration will be realized by minimizing
the correlation energy, that is, the energy associated with these clusters, 
\textbf{\ } 
\begin{equation}
\delta E_{q}=0,  \label{eqn:dener}
\end{equation}
with%
\begin{equation}
M=Const.  \label{Mconst}
\end{equation}
where the variation should be taken with the occupation numbers, $\left\{
N_{\alpha}\right\} .$ This is equivalent to the variational problem of the
following quantity, 
\begin{equation}
\delta\sum_{\alpha}\varepsilon_{\alpha}N_{\alpha}^{q}=0,  \label{deltaEq}
\end{equation}
with the following constraints, 
\begin{align}
\sum_{\alpha}N_{\alpha} & =N=Const,  \label{N} \\
\sum_{\alpha}N_{\alpha}^{q} & =A=Const.  \label{Nq}
\end{align}
That is, we consider the following variational problem, 
\begin{equation}
\delta\left[ \sum_{\alpha}\varepsilon_{\alpha}\ N_{\alpha}^{q}+\lambda
\sum_{\alpha}N_{\alpha}^{q}-\mu\sum_{\alpha}N_{\alpha}\right] =0,
\label{Variation}
\end{equation}
for all $N_{\alpha}$' s. We get%
\begin{equation}
\left( \varepsilon_{\alpha}+\lambda\right) N_{\alpha}^{q-1}-\mu =0,
\label{solN}
\end{equation}
If we introduce the occupation probability%
\begin{equation}
p_{\alpha}=\frac{N_{\alpha}}{N},  \label{pa}
\end{equation}
then the above solution can be expressed as 
\begin{equation}
p_{\alpha}=\left[ \frac{\mu^{\prime}}{\lambda q}\left( \frac{1}{%
1+\varepsilon_{\alpha}/\lambda}\right) \right] ^{\frac{1}{q-1}},  \label{min}
\end{equation}
where $\mu^{\prime}=\mu/N^{q-1}$. Calling $\lambda=\left( q-1\right) T,$
this distribution can equivalently parametrised in terms of $q$-exponential
function as,%
\begin{align}
\frac{p(E)}{p\left( 0\right) } & =\left( \frac{1}{1+\frac{1}{q-1}\frac {E}{T}%
}\right) ^{\frac{1}{q-1}}  \notag \\
& =Exp_{q}\left( -\frac{E}{T}\right) .  \label{q-Spectrum}
\end{align}
This form of parametrisation is convenient since it reduces to the the
Boltzmann-Gibbs distribution function in the limit of $q\rightarrow1$.

However, it should be stressed that the thermodynamic properties of the
prethermal state proposed here are different from that of the
Boltzmann-Gibbs statistics. In usual derivation of the Boltzmann
distribution, we have to maximise the total number of microstates accessible
for a given total energy and number of particles. The variation principle
Eq.(\ref{Variation}) in its form does not lead to the Boltzmann-Gibbs
variational equation for maximising the entropy of the system, since for $%
q\rightarrow1,$ the second term and the third term degenerates if $\lambda$
and $\mu$ are finite constant, independent of $q$.

On the other hand, we may re-arrange these Lagrange multipliers $\lambda$
and $\mu$\ in terms of new constants, $\beta$ and $\tilde{\mu}$ as 
\begin{equation}
\delta\left[ \sum_{\alpha}\varepsilon_{\alpha}\ N_{\alpha}^{q}+\frac{1}{%
\beta\left( q-1\right) }\left( N-\sum_{\alpha}N_{\alpha}^{q}\right) -\tilde{%
\mu}\sum_{\alpha}N_{\alpha}\right] =0,  \label{Variation2}
\end{equation}
where 
\begin{align}
\frac{1}{\beta\left( q-1\right) } & =\lambda \\
\tilde{\mu} & =\mu+\frac{1}{\beta\left( q-1\right) }
\end{align}
and we have used%
\begin{equation}
\sum_{\alpha}N_{\alpha}=N.
\end{equation}
The second term 
\begin{align}
\frac{1}{\beta\left( q-1\right) }\left( N-\sum_{\alpha}N_{\alpha}^{q}\right)
& =\frac{1}{\beta}\sum_{\alpha}\frac{1-N_{\alpha}^{q-1}}{q-1}N_{\alpha} 
\notag \\
& =-\frac{1}{\beta}\sum_{\alpha}\ln_{q}\left( N_{\alpha}\right) N_{\alpha}
\label{Tsallis_S}
\end{align}
where 
\begin{equation}
\ln_{q}\left( N_{\alpha}\right) \equiv\frac{1-N_{\alpha}^{q-1}}{1-q}
\label{lnq}
\end{equation}
and in the limit of $q\rightarrow1,$ we have%
\begin{equation*}
\ln_{q}\left( N_{\alpha}\right) \rightarrow\ln\left( N_{\alpha}\right) .
\end{equation*}
Thus, for $q\rightarrow1,$ the variational principle, Eq.(\ref{Variation2})
leads to%
\begin{equation}
\delta\left[ \sum_{\alpha}\varepsilon_{\alpha}\ N_{\alpha}-\frac{1}{\beta }%
\sum_{\alpha}N_{\alpha}\ln N_{\alpha}-\tilde{\mu}\sum_{\alpha}N_{\alpha }%
\right] =0,  \label{VariationBoltzmann}
\end{equation}
which is exactly the same equation we get in the Boltzmann-Gibbs statistics
when we maximise the entropy%
\begin{equation}
S_{BG}=k\ln W,
\end{equation}
with%
\begin{align}
\ln W & =\ln\left( \frac{N!}{\prod_{\alpha}N_{\alpha}!}\right) \\
& \simeq-\sum_{\alpha}N_{\alpha}\ln N_{\alpha}+const.
\end{align}
after using Stirling's formula.

All the discussion above can be done in terms of occupation probability, $%
p_{\alpha}$ from the beginning. The variational problem, Eq.(\ref{deltaEq})
together with the constraints, Eqs.(\ref{N}) and (\ref{Nq}) is equivalent to%
\begin{equation}
\delta\left( \sum_{\alpha}\varepsilon_{\alpha}p_{\alpha}^{q}\right) =0
\end{equation}
with%
\begin{equation}
\sum_{\alpha}p_{\alpha}=1
\end{equation}
and%
\begin{equation}
\sum_{\alpha}p_{\alpha}^{q}=B,
\end{equation}
where the variation is now taken with respect to $\left\{ p_{a}\right\} .$
The corresponding equation to Eq.(\ref{Variation2}) is 
\begin{equation}
\delta\left( \sum_{\alpha}\varepsilon_{\alpha}p_{\alpha}^{q}-\frac{1}{\beta }%
\sum_{\alpha}p_{\alpha}\ln_{q}\left( p_{\alpha}\right) +\mu\sum_{\alpha
}p_{\alpha}\right) =0  \label{Variation3}
\end{equation}
where the second term is nothing but Tsallis' entropy 
\begin{equation}
S_{Tsallis}=-\sum_{\alpha} p_{\alpha} \ln_{q}\left( p_{\alpha}\right) .
\label{STsallis}
\end{equation}
Thus, we have shown formally that our system is a physical realization of
Tsallis' formulation of the generalised statistics.

Here, there are important remarks. Although we have shown that the energy
minimum principle of correlated clusters lead to the same results as
Tsallis' generalised statistics, the physical significance of our example is
very different from that of Tsallis' approach. In our approach, we imposed
the condition to minimise the energy of systems of $q$-clusters, keeping the
total number of clusters. This energy is not the total energy of the system,
but that of a subsystem of the system which are strongly correlated. When we
neglect the interaction energies, the total energy of the system is the sum
of single particle energies so that 
\begin{equation}
E_{total}=\sum_{\alpha }\varepsilon _{\alpha }N_{\alpha }.  \label{Etot}
\end{equation}%
In this case, if we require conservation of the total energy of the system,
we have to add the constraint, 
\begin{equation}
\delta E_{total}=0
\end{equation}%
in addition to Eqs.(\ref{N}),(\ref{Nq}). By introducing a new Lagrange
multipliers, we obtain instead of Eq.(\ref{eqn:dener}), 
\begin{equation}
\frac{p(E)}{p\left( 0\right) }=\left( \frac{1-E/\varepsilon _{\max }}{1+%
\frac{E/T}{q-1}}\right) ^{\frac{1}{q-1}}  \label{p_consE}
\end{equation}%
where now $p\left( 0\right) ,T$ and $\varepsilon _{\max }$ should be
determined from the constraints, Eqs.(\ref{N}),(\ref{Nq}) and (\ref{Etot}).
As expected, the energy spectrum has the maximum cut-off energy, $%
\varepsilon _{\max }$ which is a natural consequence of the conservation of
energy. For $E\ll \varepsilon _{\max },$ the above expression reduces to Eq.(%
\ref{q-Spectrum}) as expected.

In general, the number of the $q$-clusters and the coefficient of the
probability of the cluster formation will depend on $\alpha$. In this case,
we may write the number of clusters as 
\begin{equation}
M_{\alpha}=K_{\alpha}\frac{N_{\alpha}^{q_{\alpha}}}{\Gamma(q_{\alpha}+1)}.
\end{equation}
The variational problem is then 
\begin{equation*}
\delta\left[ \sum_{\alpha}q_{\alpha}K_{\alpha}\varepsilon_{\alpha}\
p_{\alpha}^{q_{\alpha}}+\lambda\sum_{\alpha}K_{\alpha}p_{\alpha}^{q_{\alpha
}}-\mu\sum_{\alpha}p_{\alpha}\right] =0.
\end{equation*}
Then the distribution function is 
\begin{equation}
\frac{p(E)}{p\left( 0\right) }=\left( \frac{K_{0}/K_{E}}{1+\frac{q_{\alpha }%
}{q_{\alpha}-1}\frac{E}{T}}\right) ^{\frac{1}{q_{\alpha}-1}}.
\end{equation}

\section{Toy model of correlated particles}

To check our prethermalization scenario, let us consider a very simple toy
model described below.

\begin{enumerate}
\item Initially $N$ particles are distributed as $\left\{
N_{0},N_{1},...,N_{\alpha},...\right\} $ over equally spaced energy levels.
Let us denote the energy of the $i-$th particle as $E_{i}.$

\item Choose randomly a pair of particles, say, $i$ and $j.$

\item Energies of $i$ and $j$ are updated according to one of the following
alternatives:

\begin{enumerate}
\item The new energies are set to the lower one of $E_{i}$ and $E_{j},$ that
is,%
\begin{equation}
E_{i},E_{j}\rightarrow E_{i}^{\prime}=E_{j}^{\prime}=Min\left(
E_{i},E_{j}\right) ,
\end{equation}
then choose another $k$-th $\left( k\neq i,j\right) $ particle randomly and
attribute 
\begin{equation}
E_{k}^{\prime}=E_{k}+Max\left( E_{i},E_{j}\right) -Min(E_{i},E_{j})
\end{equation}
to conserve the total energy.

\item Change the energies as 
\begin{align}
E_{i} & \rightarrow E_{i}^{\prime}=E_{i}\pm\Delta E,  \notag \\
E_{j} & \rightarrow E_{j}^{\prime}=E_{j}\mp\Delta E,
\end{align}
where $\Delta E$ is the level spacing. Here, if one of $E_{i}^{%
\prime},E_{j}^{\prime}$ becomes negative, then this step is skipped. That
is, we have to keep always $E_{i}\geq0$.
\end{enumerate}

\item The alternatives ($a$) and ($b$) are chosen randomly, but the ratio of
the average frequency of ($a$) to ($b$) is kept as a constant, $r$.
\end{enumerate}

Here, the collision of type ($a$) is the process of the formation of a kind
of clusters, while the collision of type ($b$) is the energy exchange
process and may destroy the formed clusters. It is well-known that for $r=0$
(no type (a) collision) in the above model, the final single-particle
spectrum will become the Boltzmann distribution. By the competition of the
two collision processes, we expect the number of correlated particles will
reach some stationary value, and, by construction, the energy of a
correlated pair always tends to diminish. In this way, we may expect that
the above system will lead to the situation described by Eq.~(\ref{min}). In
Fig.~2, we show the results of simulations, for several values of $r$ from $%
0.00001$ to $0.1$. It is interesting to note that $r>0$ leads to a
non-Boltzmann distribution which is well approximated by the Tsallis
distribution, $p_{a}=C/\left[ 1+\left( q-1\right) \beta E_{\alpha }\right]
^{1/(1-q)}$, where $\beta =1/(3-2q)\left\langle E\right\rangle $ and $%
C=(2-q)/(3-2q)\left\langle E\right\rangle $ are determined from the
normalization condition and conservation of energy. One parameter fits with
respect to $q$ were performed and the results are indicated by the
continuous curves in this figure. For $r\rightarrow 0$, the spectrum
converges to the Boltzmann distribution $p_{a}\rightarrow e^{-E_{\alpha
}/\left\langle E\right\rangle }$ as expected. Note that for a one
dimensional case like the present model, the Tsallis distribution is valid
only for $q<3/2$, otherwise the energy expectation value diverges. For
larger values of $r$, the fitted value of $q$ tends to this limiting value,
but the distribution begins to deviate substantially from the Tsallis
distribution in the low energy region.

\begin{figure}[tbp]
\includegraphics[scale=0.6]{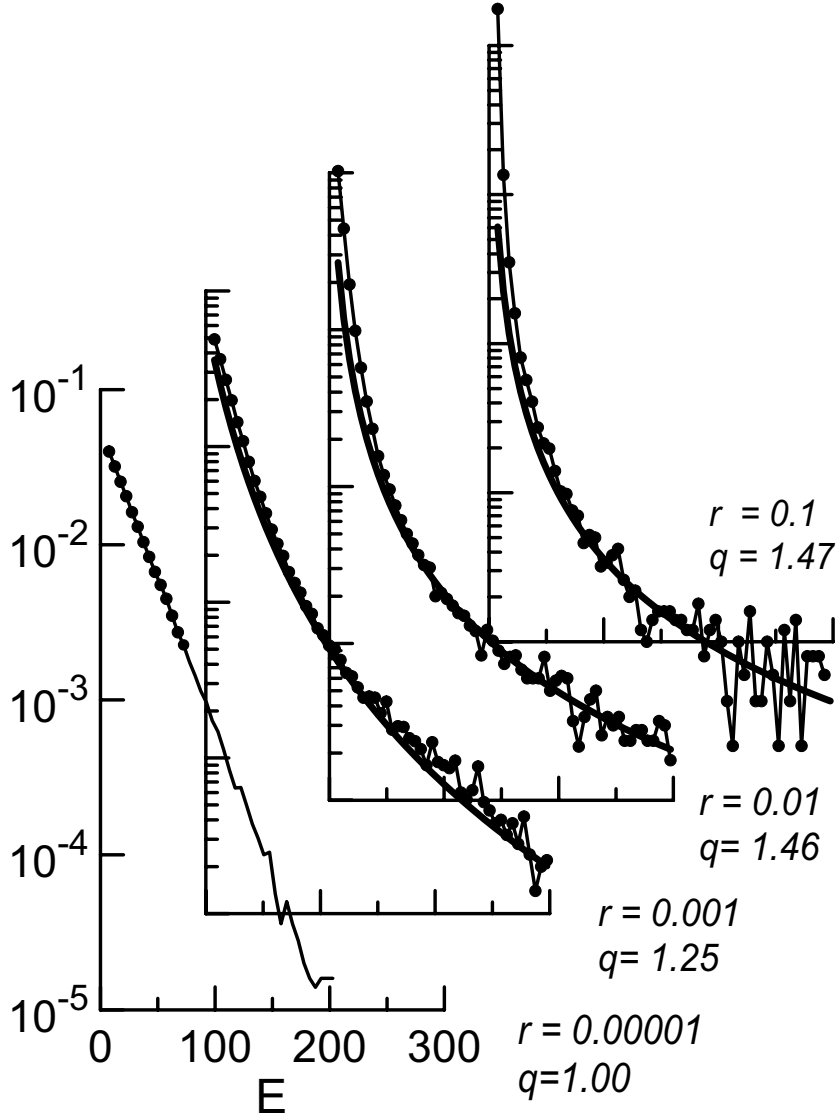}
\caption{Energy spectrum after a large number of collisions per particle,
starting from a distribution peaked at $\left\langle E\right\rangle =22.5$.
The probability of cluster-forming collisions ranges from $r=0.1$ (top) to $%
r=0.00001$ (bottom). Tsallis distributions fitted to the calculated spectra
are also shown, together with the corresponding $q$-values.}
\label{fig2}
\end{figure}

It is important to note that these distributions are the stationary and
stable ones. We confirm that starting from any different initial conditions,
the final distribution converges uniquely for a given $r$ parameter.
Furthermore, the cluster formation through type (a) collisions accelerate
the speed of the convergence. For example, for $r=0.01$ the distribution
converges 10 times faster than $r=0$ case.

In the above toy model, the time reversal is violated for the collision type
($a$). However, this is not the crucial factor to obtain the non-Boltzmann
distribution. We have checked this in a more elaborate model which has
time-reversal invariance.

\section{Interacting Linear Harmonic Oscillators}

As an another simple example in which a single particle spectrum deviates
from a simple exponential form, let us consider $N$ identical harmonic
oscillators $\vec{q}\left( t\right) $, described by the following
Hamiltonian,%
\begin{equation}
H=\frac{1}{2}\left( \frac{d\vec{q}}{dt}\right) ^{2}+\frac{1}{2}\vec{q}^{2}+%
\frac{1}{2}G\vec{q}^{T}C\vec{q}.
\end{equation}%
Here%
\begin{equation}
\vec{q}=\left( 
\begin{array}{c}
q_{1}\left( t\right) \\ 
q_{2}\left( t\right) \\ 
\vdots \\ 
q_{N}\left( t\right)%
\end{array}%
\right) ,
\end{equation}%
and $C$ is an $N\times N$ symmetric matrix, representing the linear
interactions among these $N$ harmonic oscillators. The constant $G$
represents the magnitude of coupling. It is well-known that such a system is
formally solvable. Let $U$ the orthogonal transformation which diagonalize
the matrix $C,$%
\begin{equation}
UCU^{T}=\Lambda ,
\end{equation}%
where $\Lambda $ is the diagonal matrix whose diagonal elements are
eigenvalues of $C.$%
\begin{equation}
\Lambda =diag(\lambda _{1},\lambda _{2},..,\lambda _{N}).
\end{equation}%
Then by introducing the new vector,%
\begin{equation}
\vec{\xi}\left( t\right) =U\vec{q}\left( t\right) ,
\end{equation}%
we have%
\begin{equation}
H=\frac{1}{2}\left( \frac{d\vec{\xi}}{dt}\right) ^{2}+\frac{1}{2}\vec{\xi}%
^{2}+\frac{1}{2}g\vec{\xi}^{T}\Lambda \vec{\xi}
\end{equation}%
so that the general solution for $\vec{\xi}$ is given by%
\begin{equation}
\xi _{i}\left( t\right) =\frac{1}{\omega _{i}}A_{i}\sin \left( \omega
_{i}t\right) +B_{i}\cos \left( \omega _{i}t\right) ,\ \ i=1,..,N  \label{sol}
\end{equation}%
where%
\begin{equation}
\omega _{i}=\sqrt{1+g\lambda _{i}}
\end{equation}%
and $\vec{A}=\left( A_{i}\right) $ and $\vec{B}=\left( B_{i}\right) $ can be
determined from the initial condition given for $\vec{q}$ and $d\vec{q}/dt$
as 
\begin{equation}
\vec{B}=U\vec{q}\left( 0\right)  \label{B}
\end{equation}%
and%
\begin{equation}
\vec{A}=U\frac{d\vec{q}\left( 0\right) }{dt}.  \label{A}
\end{equation}

Eq.(\ref{sol}), together with Eqs.(\ref{B}) and (\ref{A}) shows that the
whole system never realize the thermally equilibrated state, which is a very
known fact. On the other hand, a single particle trajectory in a reduced
phase space $\left( q,p\right) $ is very complicated and even ergodic if the
coupling matrix $C$ is sufficiently complex. It is then interesting to see
how the long-time average of single particle phase space distribution of any
of $q_{i}\left( t\right)$'s behaves. Let us define 
\begin{equation}
f\left( q,\dot{q}\right) \Delta q\Delta\dot{q}=\lim_{T\rightarrow\infty }%
\frac{1}{T\ }\frac{1}{N}\sum_{i}\int_{0}^{T}dt\ \Theta^{\left( 2\right) }%
\left[ \left( q_{i}\left( t\right) ,\dot{q}_{i}\left( t\right) \right)
;\Delta\Omega\right] ,
\end{equation}
where $\Theta^{\left( 2\right) }$ is $2-$dimensional Heaviside's step
function defined by%
\begin{eqnarray}
\Theta^{\left( 2\right) }\left[ \left( q_{i}\left( t\right) ,\dot{q}%
_{i}\left( t\right) \right) ;\Delta\Omega\right] = \left\{ 
\begin{array}{cc}
1 & if~\left( q_{i}\left( t\right) ,\dot{q}_{i}\left( t\right) \right)
\in\Delta \Omega , \\ 
0 & otherwise,%
\end{array}
\right.
\end{eqnarray}
and $\Delta\Omega$ is an infinitesimal domain around the phase-space point $%
\left( q,\dot{q}\right) $ whose volume is given by $\Delta q\Delta\dot{q}.$
In the above, we assume that any of harmonic oscillators are identical, so
we consider the inclusive single particle distribution of these $N$
oscillators.

\begin{figure}[tbp]
\includegraphics[scale=0.6]{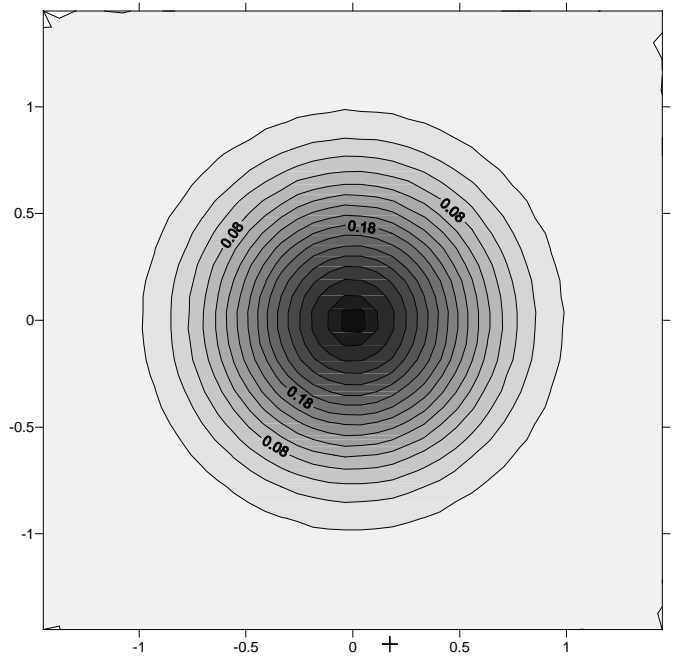}
\caption{Phase Space distribution of single particle states. The abscissa
represents the $q$ coordinate, and the ordinate represents the $\dot{q}$
coordinate.}
\label{fig3}
\end{figure}

In Fig. \ref{fig3}, we show the results of numerical calculation for the
system with $N=100$ and the symmetric matrix $C$ is created by a homonegeous
random number $C_{ij}\in \left[ -1,1\right] $. Also a random initial
condition,%
\begin{align*}
q_{i}\left( 0\right) & \in \left[ 0,1\right] , \\
\dot{q}_{i}\left( 0\right) & \in \left[ -1,1\right] ,
\end{align*}%
has been specified. The result seems an Gaussian distribution in $\left(
q,p\right) $ plane. To see more precisely, we show the average of single
particle spectra 
\begin{equation}
P\left( E\right) =\frac{1}{N}\sum_{i}P_{i}\left( E\right) ,  \label{P(E)}
\end{equation}%
where 
\begin{equation}
P_{i}\left( E\right) dE=\lim_{T\rightarrow \infty }\frac{1}{T}%
\int_{0}^{T}dt\ \Theta \left( \left( \frac{dE}{2}\right) ^{2}-\left(
E_{i}\left( t\right) -E\right) ^{2}\right) ,  \label{Pi(E)}
\end{equation}%
and $E_{i}\left( t\right) $ is the single particle energy without
interaction term, 
\begin{equation}
E_{i}=\frac{1}{2}p_{i}^{2}+\frac{1}{2}q_{i}^{2}+\frac{1}{2}gC_{ii}q_{i}^{2},
\label{SingleE}
\end{equation}%
which corresponds to the single particle oscillator energy with renormalised
frequency \footnote{%
Another possible way to define the single particle energy is to take the
part of Hamiltonian depending on $i$-th coordinate, $E_{i}=\frac{1}{2}%
p_{i}^{2}+\frac{1}{2}q_{i}^{2}+\frac{1}{2}g\sum_{j=1}^{N}C_{ij}q_{i}q_{j}$
such that $\sum_{i=1}^{N}E_{i}=E_{ToT}$. However, this energy is not
positive definite depending on the interaction.}

\begin{figure}[tbp]
\includegraphics[scale=0.6]{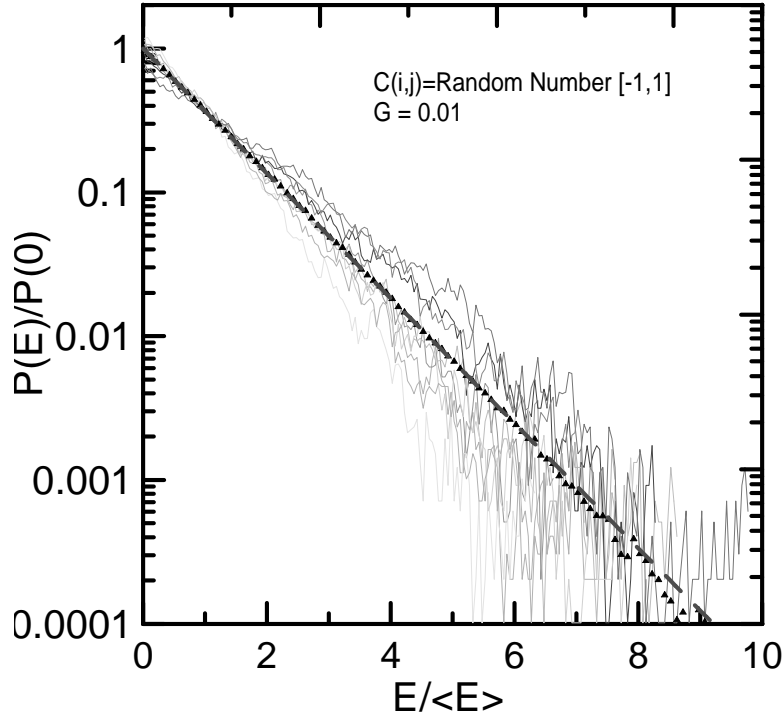}
\caption{Energy spectra corresponds to the example of Fig. \protect\ref{fig3}%
. The numerical results (the triangles) are fitted by the Boltzmann
distribution function (The dashed line). Thin lines are spectra for several
individual oscillators.}
\label{fig4}
\end{figure}

In Fig. \ref{fig4}, we show the spectrum corresponding to the example of
Fig. \ref{fig3}, which in fact, the inclusive single particle behavior is
that of the thermal equilibrium. In Fig. \ref{fig4}, the triangles are those
calculated from the expression (\ref{P(E)}), whereas the dashed line
indicate the Boltzmann distribution,%
\begin{equation}
P_{B}\left( E\right) \propto e^{-E/\left\langle E\right\rangle },
\end{equation}
where $\left\langle E\right\rangle =E_{Tot}/N.$ We see that the inclusive
energy spectrum of harmonic oscillators is essentially described by the
Boltzmann distribution. However, we should note that if we see more
carefully individual spectra for distinct oscillators, they fluctuate around
the Boltzmann distribution (thin lines).

\begin{figure}[tbp]
\includegraphics[scale=0.6]{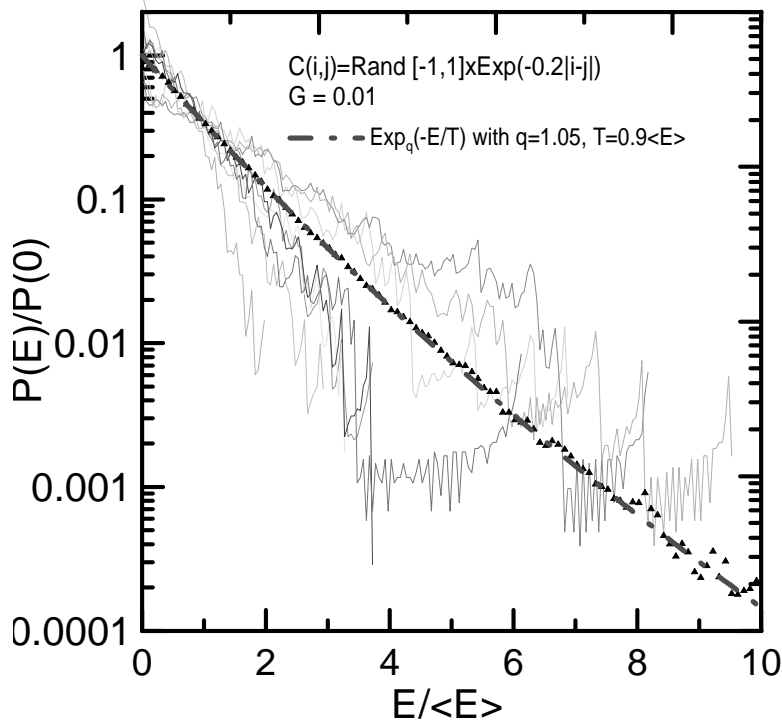}
\caption{The energy spectra in the case of the semi-diagonal random
coupling. The triangles correspond to the calculated inclusive spectrum and
dashed-dot line is the fit by Exp$_{q}$ function. Thin lines are spectra for
several individual oscillators.}
\label{fig5}
\end{figure}

Such a behavior can be seen in various initial conditions and $C$ matrix.
The above example corresponds to the situation where all the oscillators
couple each other, a typical case of long range interaction. We may consider
more local interaction, that is, the matrix is semi-diagonal. Let us
consider, for example, 
\begin{equation}
C_{ij}=\left[ R\right] e^{-\alpha \left\vert i-j\right\vert }
\label{semidiagnal}
\end{equation}%
where $\left[ R\right] $ is a uniform random number between $\left[ -1,1%
\right] $ and $\alpha $ is a parameter related to the locality of the
interaction. In Fig. \ref{fig5}, we show the inclusive spectrum for the case
of $\alpha =0.2.$ Surprisingly, the inclusive spectrum (triangles) deviates
appreciably from the expected Boltzmann distribution. Instead, the form of
the spectrum can be fitted very well in terms of $Exp_{q}\left( -E/T\right) $
function, with $q=1.05$ and $T=0.9\left\langle E\right\rangle $.

The reason why the inclusive spectrum deviates from the Boltzmann to Tsallis
($Exp_{q}$) distribution is not obvious. However, we may note several key
aspects indicating this result. First, the semi-diagonal nature of the
matrix $C$ means that there are more correlations among oscillators $\vec{q}$
than they are uniformly coupled among them. Second, the individual spectra
for different oscillators fluctuate much more compared to the case of
uniformly interacting case (Fig. \ref{fig4}). Third, the effective
temperature $T$ is smaller than the Boltzmann case. The last, but not least,
important point is that the inclusive spectrum we are looking in this
coupled harmonic oscillators does not account for the interaction energy.
That is, the sum of single particle energies defined in Eq.(\ref{SingleE})
does not give the total energy of the system. The cross terms as $%
q_{i}C_{ij}q_{j}$ for $i\neq j$ are not counted. These terms are omitted in
the definition of single particle energies, expecting that these are
\textquotedblleft irrelevant\textquotedblright\ information and cancels out
for the behavior of single particle energy in average. However, these
effects become relevant in the presence of correlations.

Of course, there are many cases where the inclusive spectrum is neither
Boltzmann nor Tsallis. For example, if we generate $C_{ij}$ as homogeneous
random numbers between $\left[ 0,1\right] $ (that is, all $C_{ij}>0$ ) or
between $\left[ -1,0\right] $ (all $C_{ij}<0$ ), the inclusive spectrum are
completely different.

It is then interesting to study the variation of the inclusive spectrum with
respect to the form of the coupling matrix $C.$ However, it is not a trivial
task to choose the coupling matrix $C$ in an arbitrary manner, since we have
to have all the eigenvalues real and positive to guarantee the stability of
the system. From this reason, we may generate the interaction matrix
starting from its eigenvalues. First we choose the set of $N$ positive
eigenvalues. Then we construct a random unitary matrix $U$ to define%
\begin{equation}
C=U^{T}\Lambda U
\end{equation}%
The generation of the random unitary matrix can be done using the Schmidt
orthonormalization method from a set of arbitrary $N$ linearly independent
vectors, $\left\{ \vec{e}_{i}\right\} .$

\begin{figure}[tbp]
\includegraphics[scale=0.6]{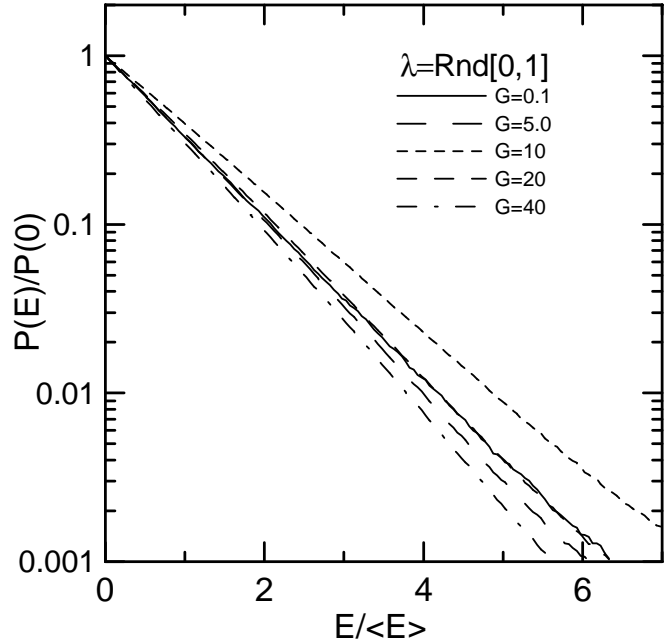}
\caption{the sequence of inclusive spectra varying the coupling constant $G$
for the eigenvalues of $C$ randomly chosen between $\left[ 0,1\right]$. }
\label{fig6}
\end{figure}

In Fig. \ref{fig6}, we show the sequence of inclusive spectra varying the
coupling constant $G$ for the eigenvalues of $C$ randomly chosen between $%
\left[ 0,1\right] $. The initial condition are taken randomly generated as
in the examples before. We can see that all the inclusive spectra are
consistent to the Boltzmann distribution up to the energy plotted. (For
larger energy, the spectra deviate from the Boltzmann due to the finiteness
of $N$ ($=100$ ).

\begin{figure}[tbp]
\includegraphics[scale=0.6]{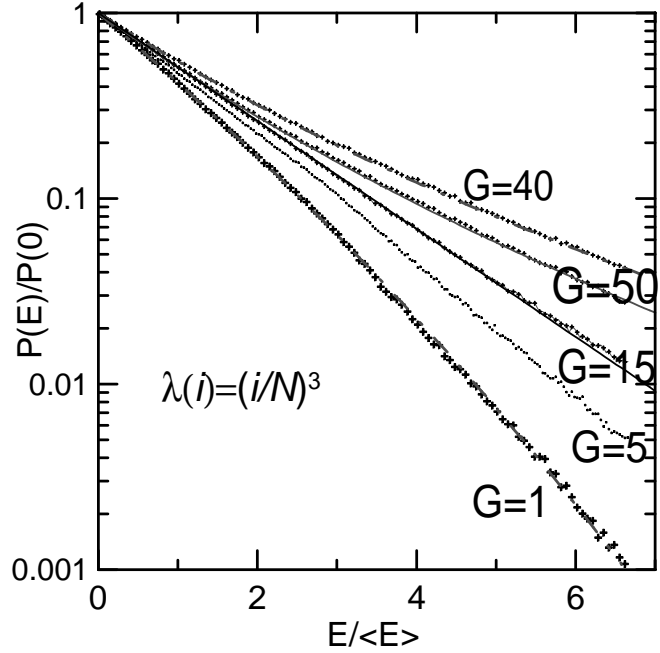}
\caption{The sequence of inclusive spectra when the eigenvalues are very
inhomogeneous.}
\label{fig7}
\end{figure}

On the other hand, in Fig. \ref{fig7}, we show the sequence of inclusive
spectra when the eigenvalues are very inhomogeneous. In this example, we
took eigenvalues as a quickly increasing function of its index, 
\begin{equation}
\lambda _{k}=\left( \frac{k}{N}\right) ^{3}.
\end{equation}%
A similar behavior of the inclusive spectrum appears when we take 
\begin{equation}
\lambda _{k}=e^{\ \zeta k/N},
\end{equation}%
where $\zeta $ is a constant. We found that all of these spectra are very
well fitted by the two parameter Tsallis distribution, $Exp_{q}\left(
-E/T\right)$. The negative curvature spectra also appear, but they can also
be fitted by $q<1$.

It is interesting to observe that other simple 2-parameter function, such as
exponential of polynomial in $E,$%
\begin{equation}
\exp \left( -aE+bE^{2}\right)
\end{equation}%
does not work at all.

Another important fact is that, as we mentioned in the introduction, the
correlation among the internal degrees of freedom may depend on the initial
condition so that the value of $q$ can also depend on the initial condition.
In fact, the form of inclusive spectrum when we change the initial condition
changes but still can be expressed by $Exp_{q}(-E/T),$ with different $q$
and $T$. Therefore, the parameter $q$ is rather a kind of state parameter
like other thermodynamic quantities and not the parameter specified by the
Hamiltonian of the system.

\section{Summary and Concluding Remarks}

In this paper, we discussed the concept of the prethermalized state as the
presence of strong correlation in the phase space of the system. We argued
that those states described by Tsallis statistics belong to this category.
We show that a very specific class of correlation leads naturally to the
Tsallis distribution for the occupation probability in energy levels.

We also show an interesting simple example where Tsallis distribution
emerges. We studied the inclusive energy spectrum of $N$ coupled harmonic
oscillators. Depending on the coupling, the inclusive spectrum exhibit
clearly the Tsallis distribution. The discussion here have been
restricted in the framework of classical mechanics. However, the
mathematical structure of general coupling matrix is equivalent to a quantum
mechanical problem to find the eigen frequencies if the vector 
$\vec{q}$ is considered as a state vector in an appropriate base. The single
particle spectra are thus somewhat related to the reduced density matrix
defined for each individual oscillators. In this sense, the question treated
here can also be related to the problem of decoherence and locality. 

The emergence of Tsallis distribution is not deterministic, but
probabilistic in the sense that it depends on the choice of the random
unitary transformation. There are specific cases that the inclusive spectrum
does not behave neither Boltzmann nor Tsallis, at all. However, when we
choose the unitary transformation randomly, then the probability of emerging
Boltzmann, or Tsallis type distribution seems to be dominant. For example, 
if we go from a system where the coupling strength is small and the
interaction matrix is uniformly random to the system where there exist
stronger interactions or correlations, the inclusive spectrum starts to
deviate from the Boltzmann to Tsallis, statistically. This observation
deserves a further systematic investigation. It will be interesting if we
can formulate the problem from the point of variational problem as discussed
in this paper.

We thank fruitful discussion with Carlos Aguiar and Gabriel Denicol.

\end{document}